# Historical Parallels between, and Modal Realism underlying Einstein and Everett Relativities

Sascha Vongehr

National Laboratory of Solid-State Microstructures, Nanjing University, Nanjing 210093, P. R. China

A century ago, "*past*" and "*future*", previously strictly apart, mixed up and merged. Temporal terminology improved. Today, not actualized quantum states, that is merely "*possible*" alternatives, objectively "*exist*" (are real) when they interfere. Again, two previously strictly immiscible realms mix. Now, modal terminology is insufficient. Both times, extreme reactions reach from rejection of the empirical science to mystic holism. This paper shows how progress started with the relativization of previously absolute terms, first through Einstein's relativity and now through Everett's relative state description, which is a modal realism. The historical parallels suggest mere relativization is insufficient. A deformation of domains occurs: The determined past and the dependent future were restricted to smaller regions of space-time. This 'light cone description' is superior to hyperspace foliations and already entails the modal realism of quantum mechanics. Moreover, an entirely new region, namely the 'absolute elsewhere' was identified. The historical precedent suggests that modal terminology may need a similar extension. Discussing the modal realistic connection underlying both relativities, Popper's proof of future indeterminism is turned to shatter the past already into many worlds/minds, thus Everett relativity is merely the Bell inequality violating correlation between those possible empirical pasts.

# 1  Introduction

Are virtual particles "real"; do alternative worlds "objectively exist"? We know the involved physics now, but there is no consensus on terminology. I believe that the improvement over pre-relativistic temporal terminology is not just a mere historical similarity. The involved relativizations are fundamentally the same modal realism. Keeping this in mind and taking the historical precedent as a guide should help the unification of the involved theories.

The present divides time into "past" and "future" (Fig. 1a). Experimental observation has proven these labels to be relative. If observer O moves relative to observer O*, some of O's future belongs to O*'s past. Doubting relativity of synchronicity and holding modern science suspect is one extreme response to this. The other extreme is to proclaim a so called block universe and that "time does not *exist*" (Barbour 2000)[1]. These opposing views go under various labels like "objectism" versus "eventism" (Maxwell 1985)[2], "presentism" and so on, which all try to make do with the traditional and therefore likely inadequate terminology. Terminology grows along with science supplying novel, not verification transcendent distinctions. Relativization is a first step: The previous division into past ($t < t_0$) and future ($t > t_0$), which was only relative to the choice of $t_0$, became also relative to the observers' velocities. This suggests different hyperspace foliations, which are different ways to slice the single, four dimensional space-time, which naïve realism supposes to exist 'out there'. However, relativization and differently cutting one single world are not the whole story. The *determined* past and the *dependent* future, previously identical to past and future, respectively (Fig. 1a), transformed into cones (Fig. 1b). Importantly, something previously unknown joined them: the "absolute elsewhere" (German original: "absolut Anderswo"), which is outside of an observer's past and future light cones. Light cones are *not* relative to velocity but to space-time events. The light cone description is therefore superior to hyperspace foliation, because it anticipates a modal realism about observers' determined pasts.

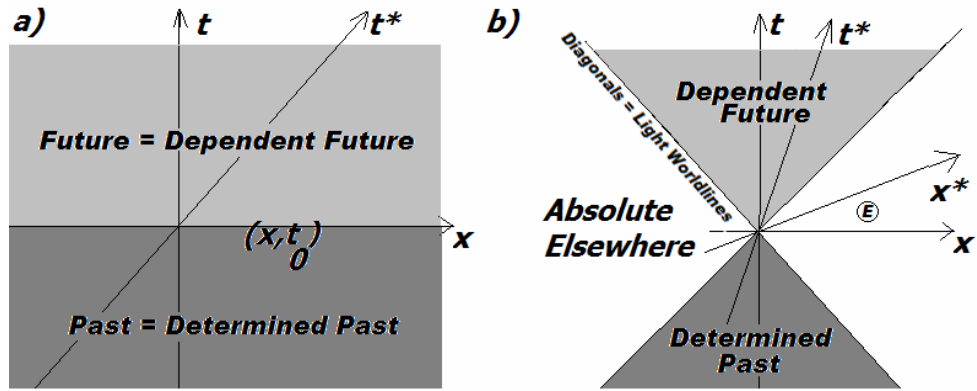

**Fig. 1** (a) Galilean relativistic Newton space-time, i.e. space, namely the *x*-direction, living through absolute time $t = t^*$: The *t*-axis is observer O's world line while the $t^*$-axis represents the history of observer O*. (b) Einstein relativistic Minkowski space-time: O* moves with about a third of the velocity of light to the right. The event E occurs at $t > t_0$ but at $t^* < t_0$, although both observers meet at $t_0$ and synchronize their clocks then. There is an entirely new region that is neither dependent future nor determined past, called the "absolute elsewhere".

The historical precedent has close parallels nowadays. There is a traditionally strict difference between what "exists" in the sense of being actualized for me here now (say I flipped a coin, it came up heads), and what was merely "possible" yet seems not actualized (it came up tails). The "possible" is either about the future (still undetermined, not yet actualized), or about potential future knowledge (a coin has already come up either head or tails, but I do not know which yet). The latter can be described as uncertainty about self-localization, about which of the possible worlds I inhabit (which, under the assumption of an extremely large or infinite universe being naively real actualized, would be in fact "localization" in the space-time terminological sense!). Traditionally, these uncertainties were very different: Something that was previously possible but now happens to have turned out otherwise, like tails when heads came up, was usually thought to be strictly non-existent, because the actual was thought to be *absolutely* actualized by the *one conserved* substance that our *single cosmos* seemed to consist of, not just *relatively* actualized to an observer. This strict differentiation is captured via three categories of modality according to Kant: "necessity" implies "existence", which in turn implies "possibility". "Possible" did *not* imply "existent". This terminology cannot consistently express whether mutually excluding possibilities in a quantum superposition "exist". States that are classically mutually exclusive, for example a fluid either spinning clockwise or counterclockwise, exist in a more concrete sense than merely potential actualizations, since they are for example simultaneously present in Schrödinger cat states (Schrödinger 1935; Lewis 2004; Wineland 2005) [3,4,5]. Superfluids can spin both ways simultaneously. In Deutsch's interpretation (Deutsch 1997)[6], quantum computing is more powerful than classical computing because the computation is distributed over all the possible 'parallel worlds'. Since they all contribute to the result, these possible worlds are "real". We can physically interact with alternative possibilities, for example rotate their quantum phase without destroying the superposition. They are physically "objective" and "exist", which makes them "real". This confirms modal realism *empirically* (Vongehr 2012)[7], also for example via Bell inequality violations by quantum experiments. Absolute actualization cannot be preserved through entangled quantum measurements; actualization rapidly spreads to multiple futures (Vongehr 2011)[8].

  The extreme reactions remind of the historical precedent. Again there are those who doubt the physics and rail against (cultural) relativism. Again there are those that indeed do carry relativism too far, claiming that 'everything possible exists' as if there is no longer anything *im*possible. As before, relativization is necessary, and the Everett relative state description (Everett 1957)[9] provided a necessary relativization of terminology without which the quantum physics cannot be coherently expressed. Everett's prose adopted relative states ontologically, but basic Everett relativity is not the same as *many worlds interpretations* (DeWitt 1973; Deutsch 1997)[10,6] or *many minds interpretations* (Albert 1988; Lockwood 1996)[11,12], which are interpretations of quantum mechanics, while basic Everett relativity can be equally applied to stochastic processes that obey classical (non-quantum) probabilities!

In the following, Section 2 analyzes the historical precedent in detail, Section 3 exhibits the contemporary parallels, and Section 4 discusses the deeper connection.

## 2   "Before" and "After" before and after Special Relativity

The present ($t_0$) divides time into the past ($t < t_0$) and the future ($t > t_0$). This is absolute in the classical, Galileo/Newtonian space-time ($x$, $t$), meaning it separates the whole space-time uniquely. We only discuss one spatial dimension $x$ for simplicity. The *present* is classically the sharp border line, the $x$-axis ($x$, $t = t_0$) that cuts the past half plane below it ($x$, $t < t_0$) from the future half plane ($x$, $t > t_0$) above (Fig.1a). Two observers O and O* move relative to each other, as is depicted by them tracing out different histories or time lines $t$ and $t^*$. In pre-relativistic physics, they agree on what constitutes the present. A separate $x^*$-axis is thus unnecessary; it equals the $x$-axis. Experimental observation has however proven that these labels are relative to the reference system: O and O* disagree on where the present cuts through space-time.

O*'s present, i.e. the $x^*$-axis ($x^*$, $t^* = t_0$), is tilted (Fig. 1b) relative to O's present and the $x^*$-axis and $t^*$-axis both lean towards the light's history (the diagonal) by the same angle. Some of O's "future" ($t > t_0$), for example the event E, belongs to O*'s "past" ($t^* < t_0$).

Extreme reactions to this failing of the strict division between past and future are to refuse relativistic physics, or to conclude determinism, the latter perhaps because it is felt that all future is already in someone's past. The first step toward a terminology capable of describing the science coherently was *relativization*. The present became a hyper surface (for example the $x$-axes) whose orientation depends on the velocity of the observer. Most problems with the relativity of synchronicity can be avoided by strictly relativistic terminology. In the famous twin paradox, twins travel far apart at high velocities. When they meet again, one is older than the other. This can be understood by merely drawing Minkowski diagrams like Fig. 1b. In hindsight, it is unsurprising that a twin has a different age after having aged differently, which in turn is unsurprising, since she traversed a different trajectory through space-time; one should be surprised if the ages still matched.

Relativization is followed and increased by a '*deformation*': The knowable or determined past (DP) is inside the past directed light cone. The dependent future (DF) is inside the future directed light cone. The rest is the absolute elsewhere (AE). The latter is comprised of all those events ($x$, $t$) at space-like (or "spatial") metric distances $ds$. "Space-like" means $(ds)^2 > 0$, with $(ds)^2 = (dx)^2 - (dt)^2$ and $dx = x - x_0$. The terminology is relativistic although velocities do no longer appear. They disappeared because "relative to velocity" equals "relative to a certain $x$-axis". The former dichotomy *(Present; Future)* was relative to reference frames like the $x^*$-axis. It is now deformed and extended into the triplet *(DP; AE; DF)*, which is relative to a space-time event ($x_0$, $t_0$). In this sense, the new terminology is even 'more relative'. The deformation into cones allows the observers O and O* (Fig. 1) to agree on their shared causal past as being the intersection of their determined pasts (see also Fig. 2). This is the seed of Everett relativity, as will be discussed in Section 4: The determined past is different to distant observers even if they are at rest relative to each other!

## 3   The Parallels with Quantum Physics

Wallace has an impressive listing (Wallace 2001)[13] of analogies between general relativity theory and quantum mechanics. It misses a few parallels that are especially relevant to our topic. Let us thus provide an independent list of similarities (S1 to 10), which focuses on the step by step progress.

### 3.1   Basics

S1 (Traditional Division):  Some traditional terminology strictly divides a certain space or set of events.  Space-time was strictly separated into two regions, past and future.  The set of all possibilities was similarly divided.  In pre-quantum physics, contra factual alternatives are strictly not actualized.  Possibilism has always criticized such actualism by suggesting relative actualization, say in completely separate universes.  However, without quantum correlations between possibilities, actualization does not necessarily spread to multiple future worlds.  Therefore, even with possibilism, as long as physics is assumed to be non-quantum, the separation into "possible" and actualized "existent" is much more rigid, and for example, as mentioned, separates different alternatives into mutually isolated universes.

S2 (Mixing): It turns out that there are overlaps between what was thought to be strictly separate. Special relativity mixed what was thought to be open future with the assumed to be already fixed past. Quantum mechanics shows that some of the previously thought merely potential, also exists. Classically mutually exclusive states interfere and thus exist in superposition inside Schrödinger states.

S3 (Infectious Mixing): The mixing is infectious: Further introductions of differently moving observers can turn more future events into past ones (Fig. 1), which threatens to render the whole future determined. In quantum physics, cautiously accepting the existence of a few select potentially actualized states implies the existence of many more. For example, assume that we accept only the simultaneous "existence" of alternative states like the dead and alive cats *as long as* they are in superposition in our laboratory, but say that we do not accept the "existence" of alternatives to ourselves. If we happen to take the dead cat out of Schrödinger's box, the alive cat does no longer "exist". However, it existed and there is no reason to presume that its consciousness just stops when we opened the box. The alive cat observes to climb out of the box alive in any case. So it still "exists" and we are forced to extend the use of that label. To stay consistent, we should now also accept that the alive cat is still in an existing superposition, but that superposition includes the alternative of us, namely those experimenters who took the alive cat out of the box. Now we are urged to extend the application of the labels even further and parallel worlds are "actualized" and "exist" although we wanted to avoid such a description by assumption.

S4 (Observer Relatedness): There is a *relativity* discovered, an observer relatedness. The fact of the present being relative to inertial systems is called *relativity of synchronicity*. The *relative state description* (Everett 1957)[9] is about that the outcome of an experiment is actualized relative to the state of the observer of that outcome. The alive cat exists relative to the experimenter that observes it. The superposition collapsed relative to those experimenters but perhaps not absolutely.

S5 (Involvement of Light): Experimental physics around the electromagnetic phenomenon of light discovered these relativities and established solid evidence for the modern perspective. In measurement theory, the more immutable a measure is, the more reliable it is. Light is in a sense *the* fundamental measure because it has no internal properties, no charges that could change and thus alter the measure. Light has no rest mass and when we asymptotically approach the light's own reference frame, all its energy will red-shift to zero, thus light does 'not exist relative to itself' in the special relativistic description. Light has itself also no time to exist (complete time dilatation) or travel path (complete Lorentz contraction). Quantum physics further reduced spin (polarization) and even the classical path of light to the mere consistency of an interaction between emitter and receiver. Of course, similar holds for all field quanta, for example electrons in the Hardy paradox setup, however only light shows that particular 'non-existence' already in special relativity. One should expect it to be of further importance when investigating how emergent "existence" can be best described as unified with the emergence of space-time and gravity from an relational perspective.

## 3.2 Controversy

S6 (Paradoxes questioning Realism): The traditional terminology leads to paradoxes: the twin paradox in relativity and the EPR paradox (Einstein 1935, Bell 1964; Aspect 1982)[14,15,16] come to mind. Paradoxes and relativity render the issues philosophically controversial. They question naïve realism, the feeling of that a particular actualization of the whole universe is 'really out there right now'. Relativity of synchronicity denied that "right now" is meaningful. Adding Everett relativity strictly disproves *local realism* (Aspect 1981)[17]. This quantum apparent non-*locality* opposes the "out there" and modifies realism anyway; merely refusing locality is not consistent, because such indirectly destroys the naïve realism that one desired to preserve. Popular science depicts the ongoing relativization as science beating philosophy, but Leibniz was more of a relativist than Einstein. On the contrary; scientists naturally defend realism as the castle they keep fortifying against irrationality. Einstein, the man associated with relativity like no other, refused to accept Everett relativity because of his brand of realism. Quantum mechanics has proven that some form of modal realism like Lewis' (Lewis 1986)[18] is necessary.

S7 (Hope for Hidden Reality): Many nevertheless cling to tradition, hoping for some loophole allowing a classical foundation after all. They often at least partially accept the experiments but reject fundamental relativity or, like Einstein, quantum indeterminacy. A hidden reality is hoped to rescue naïve realism; in relativity via a hidden Einstein-ether, in quantum physics via hidden variables. The situation today is more serious in this respect. Space-time relativity could emerge from an Einstein-Higgs-ether (Vongehr 2011)[19], for example on a string-theory like membrane universe. However, hidden variables cannot violate Bell's inequality (Bell 1966)[20], which has been clearly violated by experiments, quite recently by confirmation of the Kochen-Specker theorem (Kirchmair 2009)[21].

S8 (Holistic Over Interpretations): When the respective fields are still in their pioneering phase, even many who understand a lot of the science involved nevertheless misinterpret the discussed "Infectious Mixing" (S3) and enthusiastically try to sell Parmenides' 'all is one' description (Parmenides 1991)[22] as finally scientifically proven. Everything happens now or the 'block universe' is mixed with a principle of plenitude for actuality exclaiming "*Whatever can exist, does*" while effectively denying that anything conceivable could be impossible. Useful distinctions like possible/existent and potential/actual are lost. However, language can *make* a difference (although the difference should never be on principle verification transcendent of course). We can only ever even express, let alone discover, that vegetable and fruit are alike after having distinguished them. Identifying terms further blunts already insufficiently versatile terminology. It must be extended instead. I expand on this point also because many-world interpretations have problems that are intimately related to a blunting of modal terminology. 'Everything exists' applied to microstates that are thought to be objectively actualized out in the huge universe, results in infinite statistical ensembles which do not allow normalized probabilities. This 'measure problem' is the most serious one in modern cosmology (Page 2008)[23].

### 3.3 Progress

S9 (Rejection of Interpretation): Traditional terminology cannot describe the experimental observations coherently. Many physicists abandon it and concentrate on the mathematical formalism while refusing to be held up with interpretation. This 'shut up and calculate' attitude is known from general relativity but also at times when facing the paradoxes of special relativity, especially if Minkowski's geometrical interpretation is not employed. This attitude serves quantum chemistry and optics well. The mathematics involves concepts that are alien to everyday life, like the complex numbers due to imaginary values $ds$ from negative $(ds)^2$ that signify time-like metric distances. Imaginary numbers play a vital role in quantum mechanics, too. This adds to the mystery and difficulty in developing intuitive terminology.

S10 (Strict Relativization): Absolute terms are replaced by relative expressions. If events E and E* are at different times, "E is now relative to W" and "E* is now relative to W*" can both be true simultaneously (Saunders 1995)[24]. This goes through similarly for quantum mechanics and modal terminology. The literature mostly discusses how to *restrict* or *modify* the use of language in the context of allocation of truth values to statements made in a branching structure that offers several future possibilities [some reviewing and references in (Wallace 2005)[25]]. Any observable A has definite values only relative to other observables B. "A = $a_1$" and "A = $a_2$" cannot be both true if $a_1$ is not equal to $a_2$. "A = $a_i$ relative to B = $b_i$" can be true for all *i*.

### 3.4 Future Expectations Guiding further Progress

One should warn against overenthusiastic, artificial construction of similarities. For instance, complex numbers have disappeared from relativity theory (with few exceptions like relativistic thermodynamics). However, geometric algebra, which hides complex numbers inside rotations, and real quaternion algebra have created more confusion than insight in quantum theory, putting into doubt their claimed intuitiveness.

Nevertheless, the close parallels and connections between the temporal-relativistic case a century ago and the quantum modal-relativistic issue today suggest the following expectations (E) to guide us:

E1 (Survival): The traditional terms remain somewhat applicable and are not merely identified. Although relativity theory's entangling of space and time supports that other times *t* are no different from the present, we have not given up *making* a distinction. We still *did* something in the past; we do not only speak in the present tense. We expect that the possible/exist distinction survives its transformation.

E2 (Restriction): The traditional terms' domains transform; specifically, they are *restricted*, not just relative. The determined past was *reduced* to the events inside the past light cone. "Exist" is traditionally largely reserved for the actualized. Its future form, say "r-exist", where "r" could be read as meaning "*r*estricted", may apply only to what is directly measurable and actualized relative to the speaker. The dead and alive cats in the Schrödinger superposition are not actualized relative to us, while their whole quantum superposition is so present, but perhaps they will all be said to "r-exist". The contra factual states that are not still inside an actualized superposition, like the alive cat after we take the dead one out of the box, do no longer r-exist.

E3 (Limit Recovery): The new terminology recovers the traditional one in the classical limit. In the non-relativistic limit, which is mathematically simply gotten by letting light velocity *c* approach infinity, the determined past and dependent future coincide with the traditional past and future, respectively. The light cones (Fig. 1b) open up from 90 to 180 degrees (Fig. 1a) and the absolute elsewhere disappears. New modal terminology must be fully applicable to non-quantum cases in the precise sense that the transformed and restricted terms recover the traditional terms' domains of applicability in the classical limit, which is gotten by letting the Heisenberg constant *h* approach zero for example. The above "r-exist" would become "exist". Any entirely novel extension should disappear. Existing inside a superposition disappears in the classical limit.

E4 (Closing in on Observer): The new terminology isolates the observer, inviting charges of solipsism. This aspect goes beyond the restriction (E2). The *present* that observer O and O* can agree on when they meet is radically reduced from an infinite hyperplane (Fig. 1a) to a singular point ($x_0$, $t_0$): not only now in time but also just right there in space where the observer resides (Fig. 1b). One should expect that novel modal terminology may become extremely observer specific, maybe even subjective rather than describing what can be inter-subjectively shared even over space-time regions as small as one brain and its society of mind (Minsky 1988)[26] communicating over microseconds.

E5 (Extension): Relativization and restricting/deforming the meaning of traditional terms is insufficient. A newly discovered domain needs to be addressed. Is there a quantum analogy to the absolute elsewhere? The component states of a quantum superposition, for example the two states of Schrödinger's cat, come to mind. They are neither actualized nor contra factual alternatives. However, it is still debated whether anything counts as a strictly not realized alternative that cannot interfere anymore. 'Objective state collapse' (Penrose 1996)[27] insists on that alternatives like your eating lunch instead of reading this text now are strictly non-entangled, distinct possibilities. On the other hand, according to decoherence (Zurek 1998)[28], states are only ever For All Practical Purposes (FAPP, effectively) disentangled.

E6 (Intuitiveness): Eventually, 'shut up and calculate' may give way to quite intuitive terminology which is much less egalitarian than initially feared. In relativity, time was replaced by proper *eigen*-time, which equals physical aging and is thus a de-mystified concept. Light cone descriptions are based on physical light paths during measurements. The terminology still fails to reach common acceptance among a wide lay public, but consistent and intuitive terminology is not impossible.

## 4   Modal Realism as the Deeper Connection

Einstein and Everett relativizations are not a mere historical similarity. One can argue this in several ways: First, special relativity is more than a 'temporal modal realism'. Special relativity already shatters the classical past into a collection of possible past light cones, which each are an observer's *determined past*, even at one and the same space-time point, as we will now argue. Assuming otherwise implies emergent relativity in an Einstein-ether or a pre-determined block universe (Maxwell 1985)[29] where any stochastic behavior is divine pre-arrangement drawn on Wheeler's "great record parchment" (Wheeler 1980)[30]. A Galilean space-time is one parchment of many possible ones, and so physical law abiding quantum randomness is mysterious: Who paints the many possible parchments that way? Enter special relativity: simultaneity is relative. Therefore, if Bob's future is not determined, neither is all of his past (Fig. 2). Only the inside of his past light cone is his *determined past*. This however excludes as *un*determined some of the past light cone of spatially separated observer Alice, which is however *her determined past*: Alice's past is *un*determined although it is already determined relative to her. Hence, we must include different possible Alice*s* with their respective determined *pasts*. Popper (Popper 1956)[31] argued similarly for future indeterminism. However, taking indeterminism as self-evident from the fundamental equivalence of different future possibilities ("tautological modal realism"), the *past* parchment already disintegrates into all possible light cones! Indeterminism demands this fracturing into 'worlds' that contain the *determined empirical records* of possible observers, even if it comes with classical instead of quantum probabilities, so this is a pre-quantum conclusion.

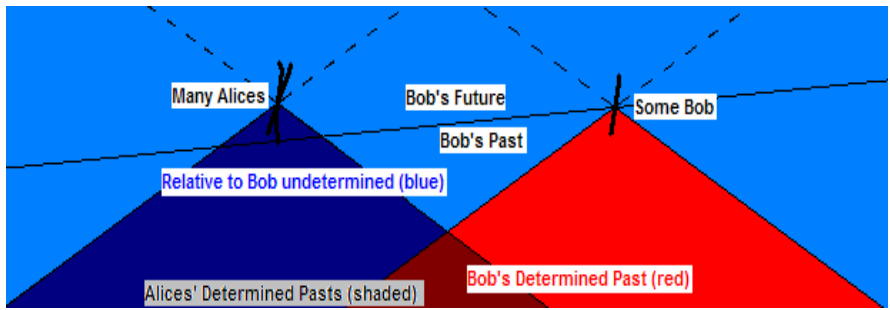

**Fig. 2** Minkowski diagram illustrating how indeterminism about the future demands different determined pasts of Alice (her parallel worlds) relative to any single determined past (any particular world) of Bob.

In other words: Everett relativity is highly suspect *without* relativity theory as the preceding step (not *with* relativity, like the EPR discussion initially suggested). For example, a non-relativistic many world universe model must quantum split everywhere into many different ones all the time. However, special relativity already deconstructed the world into an ensemble of different observers' past light cones. Therefore, merely apparent 'splitting' needs to only occur at the observation events. Everything outside of one's determined past light cone stays undetermined and does not split.

The second, more general way to argue that Einstein and Everett relativizations are not a mere historical similarity, is to notice that modern physics progressively takes the observer's situation ever more closely into account (See also E4). Special relativity looks at how the observer moves and her *knowable* past. Quantum mechanics takes into account how observation interacts with the observed and further questions, via uncertainty relations for example, what is knowable on principle. Physics leaves ontology and focuses on epistemology and phenomenology because it wants to eventually account for all that which we can be consistently conscious of, however much measurement apparatus and scientific method intervene. Via operational measurement theory, physics becomes more 'subjective', pulling back onto the observer. However, this does not imply magic quantum consciousness proposals (nor ontological commitment toward countable parallel worlds). Rather, it is a pullback onto the *describer* rather than the observer, and physics understood as the fundamental description of all that is possible implies modal realism. It is that modal realism that underlies our uncertainty and the related indeterminisms. It is likely that without modified modal terminology, neither the further fundamental uncertainty implied by quantum gravity[a] nor 'further facts' modifications of quantum mechanics, say Bell's fifth position (Gill 2002)[32] or those potentially involving consistency of phenomenal consciousness, can be understood.

---

[a] Black hole complementarity relates to further uncertainty involving the appearance of event horizons around large energies that are necessary for high resolution observation. Eventually, all uncertainties must account for the total potentiality surrounding the phenomenal end observer.

## 5  Conclusion

There is extensive philosophical work on modifying modal terminology (Kripke 1981; Lewis 1975; Stalnaker 1999; Prior 1957)[33,34,35,36] and work on category mistakes that introduced multiple senses of "exist" (Ryle 1949)[37]. Some take the input of cutting edge quantum theory very seriously into account (Saunders 1995, Wallace 2005)[24,25]. The detailed analysis of the historical precedent and its parallels as well as the deeper connection via modal realism indicate that further work along those lines, namely improving and extending modal terminology, is necessary or will anyway accompany progress on the understanding and interpretation of modern physics and the nature of reality. The analysis suggested several guidelines to keep in mind, some quite practical, like identification of a quantum analogue of the "absolute elsewhere". The most general conclusion is that everything points toward a sort of many minds modal realism, namely totality as a set of all possible observed situations as a natural starting point which leads to the Einstein and Everett relativities rather than following from them. In that view, Einstein relativity confirms modal realism, while quantum mechanics adds stricter than classically possible (common cause) correlations between the different possible worlds or minds. I suspect this to be crucial to understanding quantum foundations, namely quantum physics as the necessary correlations between alternative possibilities, including interference, superposition, and the matching of the many different Alices with the many different Bobs (Fig. 2). The self-consistency of such a many mind structure should give rise to 'reasonable' stochastic laws, the origin of which is otherwise mysterious, calling not only for one deity playing dice, but an infinite regress, every further higher-up deity throwing dice in order to provide *fair* dice to all the gods below.

# 6 References

[1] Julian Barbour: "The End Of Time: The Next Revolution In Physics." Oxford Univ. Pr. (2000)

[2] Nicholas Maxwell: "Are Probabilism and Special Relativity Incompatible?" Philosophy of Science **52**, 23-43 (1985)

[3] Schrödinger, Erwin: "Die gegenwaertige Situation in der Quantenmechanik." *Naturwissenschaften* **23**:807-849 (1935), English translation in "Quantum Theory and Measurement." Ed. J. A. Wheeler and W. H. Zurek, Princeton Univ Press (1983)

[4] Lewis, David K.: "How many lives has Schrödinger's cat?" in *Lewisian Themes: The Philosophy of David K. Lewis*. F. Jackson and G. Priest, eds., Oxford: Oxford University Press. (2004)

[5] Wineland, D. J.: "Creation of a six atom 'Schrödinger cat' state." *Nature* **Dec1**:639-642 (2005)

[6] Deutsch, David: "The Fabric of Reality." Allen Lane: New York (1997)

[7] Vongehr, Sascha: "Quantum Randi Challenge." arXiv:1207.5294v3 (2012)

[8] Vongehr, Sascha: "Many Worlds Model resolving the Einstein Podolsky Rosen paradox via a Direct Realism to Modal Realism Transition that preserves Einstein Locality." arXiv:1108.1674 (2011)

[9] Everett, Hugh: "'Relative State' Formulation of Quantum Mechanics." Rev Mod Phys **29**, 454-462 (1957)

[10] DeWitt, B. S., Graham, N.: "The Many Worlds Interpretation of Quantum Mechanics." Princeton University Press, Princeton NJ (1973)

[11] D. Z. Albert, B. Loewer: "Interpreting the many-worlds interpretation." Synthese **77**, 195-213 (1988)

[12] M. Lockwood: ' "Many minds" interpretations of quantum mechanics.' Brit. J. Phil. Sci. **47**(2), 159-188 (1996)

[13] Wallace, David: "Worlds in the Everett interpretation." Studies Hist. and Phil. Mod. Phys. **33**, 637-661 (2002)

[14] A. Einstein, B. Podolsky, N. Rosen: *Can Quantum-Mechanical Description of Physical Reality be Considered Complete?* Phys. Rev. **47**, 777-780 (1935)

[15] Bell, J. S.: "On the Einstein Podolsky Rosen paradox." Physics **1(3)**, 195–200 (1964); reprinted in Bell, J. S. "Speakable and Unspeakable in Quantum Mechanics." 2$^{nd}$ ed., Cambridge: Cambridge University Press, 2004; S. M. Blinder: "Introduction to Quantum Mechanics." Amsterdam: Elsevier, 272-277 (2004)

[16] Aspect, A., Grangier, P., Roger, G. (1982) "Experimental realization of Einstein-Podolsky-Rosen-Bohm Gedankenexperiment: a new violation of Bell's inequalities." Phys. Rev. Lett. **48**, 91-94 (1982)

[17] Aspect, A., P. Grangier, G. Roger: "Experimental Tests of Realistic Local Theories via Bell's Theorem." Phys. Rev. Letters **47**, 460-63 (1981)

[18] Lewis, David Kellogg: "On the Plurality of Worlds." Blackwell (1986)

[19] Vongehr, S.: "Metric Expansion from Microscopic Dynamics in an Inhomogeneous Universe." Com Theo Phys **54**(3), 477-483 (2011)

[20] Bell, J. S.: "On the problem of hidden variables in quantum mechanics." Rev. Mod. Phys. **38**, 447–452 (1966)

[21] G. Kirchmair, F. Zähringer, R. Gerritsma, M. Kleinmann, O. Gühne, A. Cabello, R. Blatt, and C. F. Roos: "State-independent experimental test of quantum contextuality." Nature **460**, 494-497 (2009)

[22] Parmenides of Elea, e.g.: D. Gallop: Parmenides of Elea. University of Toronto Press (1991)

[23] Page, Don N.: "Cosmological measures without volume weighting." J. Cosm. Astroparticle Phys. **10**,025, 1-24 (2008)

[24] Simon Saunders: "Time, Quantum Mechanics and Decoherence." Synthese **102**, 235-266 (1995); S. Saunders and D. Wallace: "Branching and Uncertainty." Brit. J. Phil. Sci. **59**, 293-305 (2008)

[25] D. Wallace: "Language use in a branching Universe." philsci-archive.pitt.edu/archive/00002554 (2005); some of it also in D. Wallace: "Epistemology quantized: circumstances in which we should come to believe in the Everett interpretation." Brit. J. Phil. Sci. **56**, 655-89 (2006)

[26] Minsky, Marvin: "The Society of Mind." Simon and Schuster, New York (1988)

[27] Penrose, Roger. 1996. On gravity's role in quantum state reduction. *General relativity and gravitation*. **28**(5): 581-600.

[28] W. H. Zurek: "Decoherence, Einselection and the Existential Interpretation." Phil Trans of the R Soc London A**356**, 1793–1820 (1998)

[29] N. Maxwell: "Are Probabilism and Special Relativity Incompatible?" Philosophy of Science **52**, 23-43 (1985)

[30] J.A. Wheeler: *Delayed-Choice Experiments and the Bohr-Einstein Dialogue*, The American Philosophical Society and the Royal Society, pp. 25-37, (1980)

[31] K. Popper: *The Open Universe: An Argument for Indeterminism*, Roman and Littlefield, Lanham, MD (1956)

[32] Gill, R.D.: "Time, Finite Statistics, and Bell's Fifth Position." In *Proc of "Foundation of Probability and Physics – 2" Ser. Math. Modelling in Phys., Engin. And Cogn. Sci*, **5**, (2002) pp. 179-206. Vaxjo Univ. Press, (2003)

[33] Kripke, S. A.: "Naming and Necessity." (revised and enlarged edition) Oxford: Blackwell (1981)

[34] Lewis, D.: "Languages and language." In K. Gunderson (Ed.), Minnesota Studies in the Philosophy of Science 7. Minnesota: Minnesota University Press. (1975) Reprinted in David Lewis, Philosophical Papers 1 (Oxford University Press, Oxford, 1983)

[35] Stalnaker, R. C.: "Context and Content: Essays on intentionality in speech and thought." Oxford: Oxford University Press (1999)

[36] Prior, A. N.: "Time and Modality." Oxford: Clarendon Press. (1957)

[37] G. Ryle: "The Concept of Mind." Hutchinson's, London (1949)